\documentclass[twocolumn]{article}
\usepackage{epsfig,amssymb,amsmath, cmbright}
\usepackage{amsmath}
\newtheorem{definition}{Definition}[section]
\newtheorem{theorem}{Theorem}[section]
\newtheorem{corollary}{Corollary}[section]
\newtheorem{proposition}{Proposition}[section]
\newcommand{\dom}{{\rm dom}}
\newcommand{\ch}{{\rm ch}}

\usepackage{graphics}
\usepackage{graphicx}

\usepackage{amssymb}

\begin{document}

\title{\begin{flushleft}{\bf The concept of strong and weak virtual reality\hfill\\[-.2 cm]}\end{flushleft}}
\date{\begin{flushleft}{Andreas Martin Lisewski\footnote{Email:{\tt lisewski@bcm.edu}}\\
{\scriptsize Department of Molecular and Human Genetics, Baylor College of Medicine, One Baylor Plaza, Houston, TX 77030, USA}\\{\small 31 March 2006}}\end{flushleft}}
\author{}
\maketitle


\begin{abstract}
{\begin{flushleft}\small We approach the virtual reality phenomenon by studying its relationship to set theory. This relationship provides a characterization of virtual reality in set theoretic terms, and we investigate the case where this is done using the wellfoundedness property of sets. Our hypothesis is that non-wellfounded sets (so-called hypersets) give rise to a different quality of virtual reality than do familiar wellfounded sets. We initially provide an alternative approach to virtual reality based on Sommerhoff's idea of first and second order self-awareness, and then introduce a representation of first and second order self-awareness through sets, assuming that these sets, which we call events, originally form a collection of wellfounded sets. Consequently, strong virtual reality characterizes virtual reality environments which have the limited capacity to create only events associated with wellfounded sets. In contrast, the more general concept of weak virtual  reality characterizes collections of virtual reality mediated events altogether forming an entirety larger than any collection of wellfounded sets. Aczel's hyperset theory indicates that this definition is not empty, because hypersets encompass wellfounded sets already. Moreover, we specifically argue that weak virtual reality could be realized in human history through continued progress in computer technology. Finally, we formulate a more general framework, and use Baltag's Structural Theory of Sets (STS) to show that within this general hyperset theory Sommerhoff's first and second order self-awareness as well as both concepts of virtual reality admit a self-consistent representation. Several examples and heuristic arguments are given. 
\end{flushleft}} 
\end{abstract}

\setcounter{section}{0}
\section*{\label{Introduction}Introduction}

Virtual reality has become a popular metaphor for a variety of aspects in contemporary media culture, including technological, scientific, economic, philosophic and even religious aspects \cite{rhe1991, hei1993}. The technological aspect, however, is genuine because an attempt to create virtual reality is usually seen as the output of some directed engineering process.\footnote{Here, we will not discuss possible conditions and effects as induced, for example, through intoxication, abrosia or meditation.} In recent years, noticeable technological progress has been made in this field, mainly driven by the high rate at which modern computer technology has evolved. Computer hardware, software and human-computer interfaces are nowadays the leading advances behind this phenomenon \cite{rhe1991, vrinterface}, and within a relatively short period of time they have reached a standard high enough to make virtual reality a popular technology. For example in the entertainment industry, where computer generated games have arrived at a level of sophistication in terms of their visual, acoustic and even mechanical expression unthinkable only a few decades ago \cite{videogames}. Additionally, the fast growth of the Internet has lead to a new form of virtual reality devices, often called electronic communities or virtual playgrounds \cite{second,there}, in which many human operators can participate simultaneously in a digitally designed and interactive environment provided by computer networks \cite{escape}. Apart from entertainment and recreation, virtual reality technology has found applications in military technology \cite{military1, military2}, in medicine \cite{medicine}, or in architectural design \cite{architecture} and engineering \cite{engineering}. 

Virtual reality devices can be characterized as those which through adapted technology establish an interaction with the human senses \cite{rhe1991}, e.g. with the visual, acoustic, mechanical and olfactory senses, and the quality of a realized virtual reality environment then depends on how good, or how bad, this interaction is accomplished on a technical level. In order to systematically describe this interaction, three essential quality indicators of virtual reality have been proposed \cite{ste92, hei1993, vrdef}: (a) {\it Presence}. Presence is the sense of physically being in an environment. It can be thought of as the experience of one's physical environment; it does not refer to one's surroundings as they exist in the physical world, but to the perception of those surroundings as mediated by both automatic and controlled mental processes. (b) {\it Immersion (or, vividness)}. Immersion means the representational richness of a mediated environment as defined by formal features, that is, the way which an environment presents information to the senses. (c) {\it Interactivity}. This quality refers to the degree to which users of virtual reality medium can influence the form or the content of the mediated environment. 

Any high quality virtual reality has to meet these quality indicators, and so we surmise that the essence of virtual reality is closely related to each one them. And even though virtual reality is often characterized as a phenomenon contained in computer technology, this relation reveals that it has to comply with our {\it cognitive} ability to consciously perceive  both ourselves and the external world that surrounds us.

Since the ongoing technological development in human history resembles an evolutionary process, we can ask what developmental stages virtual reality technology could reach. For instance, are there possible stages of convergence? Convergence in this context would mean that at some (future) time any further progress of virtual reality technology would not lead to a significant improvement in its quality. At present times we could still be far from such a stage, but we should acknowledge that the rate at which new developments of virtual reality related technology are presented has accelerated since the introduction of computers. This situation exemplifies that even though computer-generated virtual reality still is a recent cultural achievement, and therefore we may still qualify it as primitive, we should nevertheless recognize it as a rapidly emerging technology which possibly holds a potential to influence and to change human life and culture effectively \cite{lem1974, rhe1991, hei1993}.

Starting from these preliminary considerations, our aim is to establish and to discuss a relationship between a distinct model of human consciousness and set theory in order to extend our understanding of virtual reality as a technological  and cultural phenomenon. In particular, we want to show that virtual reality, consciousness and so-called non-wellfounded set theory are intimately related. 

Non-wellfounded set theories naturally enlarge classical, i.e. well-founded,  set theory in that they introduce new structures, called hypersets (also referred to as non-wellfounded sets in mathematical terms) which, due to their circular membership structure, cannot be represented in any conventional set theory, e.g. in the common Zermelo-Fraenkel-Axiom of Choice (ZFC) set theory. In contrast to wellfounded sets, hypersets can be thought of collections containing an {\it infinite} hierarchy of membership. Specifically, we may depict a hyperset $a$ by an infinite sequence of symbols $\{a \ni b \ni c \ni \ldots\}$, or by a circular sequence $\{a \ni b \ni c \ni a \}$. In wellfounded set theories, like in ZFC set theory, such structures are excluded by the Axiom of Foundation, which is one of the Zermelo-Fraenkel axioms stating that the membership hierarchy must be finite. We see that hypersets use the membership relation $\in$ in a circular manner---a defining character that makes hypersets well suited to analyze circular structures or recursive situations. 

As an example illustrating the difference between wellfounded and non-well\-founded objects, consider the case of bibliographical references in publications, such as scholarly articles. Normally, if a bibliographical reference is made in an article $a_3$ to a second publication $a_2$, then it follows that $a_2$ was published {\it before} $a_3$. In turn, the article $a_2$ could also refer to a third text, $a_1$, this one published even before $a_2$. This situation can be given  by a nested membership relation between three wellfounded sets: $a_3 \ni a_2 \ni a_1$. However, through widespread communication over the Internet and document storage and retrieval on the World Wide Web this normal chronological order, i.e. the wellfoundedness of bibliographical references, could be distorted. So-called Internet preprint servers nowadays allow authors to exchange manuscripts prior to editorial review and publication, so that a circular situation can form where article $a_1$ itself contains a bibliographical note pointing back to $a_3$, for instance. As a consequence, a publication of all three manuscripts would lead to a circular membership, $a_3 \ni a_2 \ni a_1 \ni a_3$, i.e a non-wellfounded structure.\footnote{This example was taken from a recent study by R. Rousseau and M. Thelwallon \cite{rou2004}, where circular arrangements of {\it hyperlinks} between {\it hypertexts} were frequently found on the World Wide Web.}  

Hyperset theories have been formulated on firm mathematical foundations \cite{bernays, acz88, dev92}, and hypersets themselves have been successfully applied in fields such as computer science, linguistics and philosophy \cite{bar1996}. But so far they have not yet been identified as indispensable structures in our capacity to consciously experience the physical world. However, we feel that precluding hypersets as elements of conscious experience {\it including} physical experience arbitrarily confines its meaning by suppressing contingent effects that virtual reality technology might have on conscious experience in the first place. One of our goals is therefore to investigate the idea of non-wellfounded objects realized in virtual reality environments, and in this context we will argue that computer technology may indeed deliver such environments. But there appear also formal reasons why hypersets may become useful for studying the interplay between human consciousness and virtual reality. As will be demonstrated in the last section of this work, two elementary qualities of consciousness can be represented in terms of modal logic, i.e. in terms of a non-classical logic that introduces modalities of propositions such as necessity and possibility. In turn, through the mathematical results of A. Baltag, modal logic can be applied readily to construct a general set theory, referred to as the Structural Theory of Sets \cite{bal99}, which naturally encompasses both sets {\it and} hypersets, thus generalizing classical set theories such as ZFC. In this approach, we address the consciousness problem by employing a model developed by G. Sommerhoff \cite{som90}, in which the `elementary qualities' in question become the categories of {\it first and second order self-awareness} defined in terms of so-called {\it internal representations} \cite{som90, som94}. These definitions put us in the position to draw a formal relationship between hyperset theory and consciousness. 

With such a relationship at hand, we may also describe reality as a collection of internal representations of the surrounding physical world, where every representation is mapped onto a set. Such mapping lends itself to classify the elements of conscious experience, which we call events, depending on the presence or the absence of wellfoundedness. That is, an event may either be associated with a wellfounded set or with a non-wellfounded set (hyperset). To be meaningful, however, such a classification of conscious experience in terms of sets must be given in a plausible and non-arbitrary manner.

We address this problem by putting forward the hypothesis that events are associated exclusively with wellfounded sets {\it originally}. Our premise is that conscious experience of the physical world during history {\it before} the development of virtual reality technology is adequately described by wellfounded sets. This is understood as reference to times in human history when the {\it cognitive picture} of the physical world was not significantly influenced or altered by technical devices, since at those times technology itself had not grown enough to exert such an influence. Only quite recently virtual reality technology has began to realize its potential and to interact with the elements of conscious physical experience. Rheingold \cite{rhe1991} gives numerous historical references to identify the original beginning and the following growth of what he calls a symbiosis between virtual reality technology and human culture. He concludes that the development of virtual reality technology has been intertwined with the evolution of human culture for many thousand years, thus dating back to the very beginnings of human expression through arts, entertainment and religion. This concatenation raises the possibility that virtual reality technology is immanently human \cite{rhe1991, hei1993}, in which case it would be difficult to identify any epoch when  virtual reality technology was  nonexistent. We acknowledge this alternative and require that the interaction at those ages in history should be small enough in terms of widespread implementation of virtual reality technology in human society. As indicated earlier, during the second half of the twentieth and throughout the beginning of the twenty-first century implementation of virtual reality technology has been driven by the rapid development of computers. We therefore refer to times before this recent development to indicate the original status when events were associated with wellfounded sets.

This premise given, we are proposing that the interaction of conscious experience with virtual reality technology may happen in two distinct modes. In a first mode, which we call strong virtual reality,  virtual reality devices may generate events that during technological evolution reach higher technical levels and thus simulate with increasing quality events associated with the physical world. For example, we may think of contemporary technology such as head-mounted-displays and data-gloves which visually and mechanically simulate natural environments as parts of the physical world. Although such devices may reach high standards, they are limited to create events represented by wellfounded sets, and so they always share--albeit in an abstract sense--a quality with the events of the original physical environment which they actually simulate. Despite its recent progress, we argue that the {\it status quo} of virtual reality technology still can be characterized as strong virtual reality. However, rapid technological advance opens another possibility, where this characterization may change to a second mode---weak virtual reality. 

In weak virtual reality, modern technical devices such as computers additionally create a new class of events, the latter which are mapped onto non-wellfounded sets. Contrary to the original events associated with wellfounded sets, those events cease to have equal natural counterparts in the original physical world. Mathematically, this situation is expressed by the fact that the totality (that is the set-universe or, simply, the universe) of wellfounded sets is a proper subclass of the totality of all hypersets (called the hyperuniverse). This idea suggests that virtual reality may eventually confront us with an unprecedented quality of human experience. Our aim is to conceptually motivate, to introduce and to explain both modes of virtual reality.

This paper is organized as follows. In the first two sections, Section \ref{reality} and Section \ref{strong}, we conceptually prepare and subsequently define strong virtual reality. As indicated, this concept employs a description of consciousness based on Sommerhoff's approach to the consciousness problem \cite{som90}. Strong virtual reality involves a set-theoretic description of what Sommerhoff calls first and second order self-awareness \cite{som90,som94}. According to Sommerhoff, these two categories are introduced as internal subjective representations necessary for conscious experience. Strong virtual reality presumes that the structure of our experience, i.e. the structure of all events represented by first order self-awareness,  becomes a universe for  Zermelo-Fraenkel-Axiom of Choice set theory. This implies that any experience we could gain from a strong virtual reality always is consistent with ZFC, or, in simpler terms, it means that any strong virtual reality cannot be the source of any experience structurally richer than the original physical world.

The next section, Section \ref{weaksub}, introduces weak virtual reality. It structurally extends the first concept in that now the totality of events in virtual reality becomes larger than any collection of events created with strong virtual reality.  A conscious subject having an impression of weak virtual reality would therefore perceive the latter as a structure {\it richer} than her original reality associated with ZFC set theory. The remainder of this section is dedicated to the question whether this definition is meaningful. First, this question is discussed theoretically by giving reference to Aczel's construction of a non-wellfounded set theory. The latter, referred to as ZFC$^-$+AFA set theory, is realized through the Anti-Foundation-Axiom (AFA) which replaces the Axiom of Foundation in Zermelo-Fraenkel set theory. Then Aczel's relative consistency result guarantees that there is a set-theoretical embedding of any universe for ZF (or for ZFC) within a larger universe for ZFC$^-$+AFA. Thus, with the help of Aczel's construction of a universe for ZFC$^-$+AFA, we obtain a candidate for the totality of events in weak virtual reality. But without further assumptions about first and second order self-awareness a conclusion on the meaningfulness of our definition cannot be drawn immediately.  We address this point again in the forthcoming section, Section \ref{empiric}, but this time giving heuristic arguments why eventually the development of computer technology may lead to weak virtual reality, and by discussing an example motivated by the artistic work of M.C. Escher.

In Section \ref{sts} we return to a theoretical consideration of the relationship between Sommerhoff's approach to consciousness and non-wellfounded set theory. We introduce some ideas of Baltag's Structural Theory of Sets (STS) \cite{bal99}, a general non-wellfounded set theory which includes Aczel's ZFC$^-$+AFA as a special case. This set theory is used as a framework for studying the above relationship, and we demonstrate why STS is a proper mathematical tool for studying the very concept of strong and weak virtual reality on the basis of Sommerhoff's ideas. This approach bridges cognitive and mathematical aspects, and so it conveys a more coherent picture of strong and weak virtual reality in terms of non-wellfounded set theory. 

\section{\label{reality} Reality, virtual reality and consciousness}

Our immediate goal is to provide an abstract but, for our purposes, useful description of reality itself. We will then use this description in order to formulate a concept of virtual reality that takes into account a consciousness model developed by Sommerhoff.

We describe reality a class $R_0$, equipped with a variety of conscious subjects each being a subclass of $R_0$. Without loss of generality, we can for now assume that there is only one such conscious subject $S$. The environment of a conscious subject is its class complement which we call the {\it world} of $S$, and we denote it as $W_0 = R_0 \backslash S$. By the {\it self} of an conscious subject we again mean the class $S$. We observe that the self and its environment are not introduced as separated entities that exist independently of each other,  but rather as mutually determining and complementary classes of a unified reality: $S$ does not exist without $W_0$, and vice versa. Moreover, we expect this mutual determinism to be essential for any proper understanding of the term `conscious', which so far has been used only as an unspecified attribute of the subject $S$. This understanding should eventually allow us to point out a relationship between the faculty of consciousness, hyperset theory and virtual reality, as mentioned in the introduction section. But as there exists a vast variety of approaches to the consciousness problem, how can we choose one that would help to meet our goals?

To make a choice, we initially take a moderate position and look for a characterization of consciousness being semantically broad enough to encompass the three constitutive qualities of virtual reality, i.e. presence, immersion and interactivity with the world. We thus do not require such a characterization  to provide us immediately with a relationship to sets or to hypersets. Instead, it should at least recognize the mutual determinism  between the conscious subject and the surrounding world in terms of these three qualities. Since all three obviously require higher cognitive functions, such a characterization is difficult with so-called `bottom-up' approaches to consciousness. Typically, a bottom-up approach initially focuses on the properties of individual brain cells, their reactions and interactions, their biochemical or biophysical properties, or their information processing capabilities. Higher cognitive functions are here expected to be detectable only {\it after} studying the concerted action of many brain cells, such as neurons. Hence, a bottom-up approach to consciousness usually does not identify higher cognitive functions in the first place, and it may even be questionable whether it can reach this goal without having a detailed definition of higher cognitive functions at hand \cite{hil93}. On the other side, a `top-down' approach acknowledges from the beginning the existence of certain higher cognitive functions necessary for conscious experience. Moreover, these functions are formulated as indispensable characteristics of consciousness seen as a {\it whole system}, which encompasses the subject as well as the subject's environment. After defining these characteristics, a top-down approach would investigate the question of how they are realized in organisms that experience consciousness, and thus it prepares an exploration of the living correlates of conscious experience \cite{hil93}. 

A `top-down' approach has been given by Sommerhoff in his book ``Life, Brain, and Consciousness''\cite{som90}. His work's conceptual basis is the identification of the subject's internal representations which he understands as the primary higher cognitive functions of consciousness \cite{som90, hil93, som94}. Our intention is to use these representations for a definition of virtual reality which explicitly accounts for our cognitive ability to be, to perceive and to interact with and within the world. Hence, we adopt Sommerhoff's systemic description of consciousness (\cite{som90}, page 90) by requiring the following two categories of representations, i.e. higher cognitive functions.
\begin{enumerate}
\item     {\it First order self-awareness}. A comprehensive and coherent internal representation of the world and the {\it self-in-the-world}.  Representations of this kind we also call events.
\item     {\it Second order self-awareness}. Representations which represent the occurrence of a representation of the first category as being part of the current state of the self.
\end{enumerate}
 
With the words of Sommerhoff, then, the faculty of consciousness is described as {\it a power of a subject to form internal representations of
category (1) and of category (2)} (\cite{som90}, page 91).
 
We are going to give three immediate comments on these categories. First, since this characterization of consciousness requires the faculty of a conscious subject $S$ to generate some {\it internal representations}, it presupposes that $S$ possesses an internal structure being able to constantly monitor the world and to register the occurrence of events. No further statements about the nature of such an internal structure responsible for this ability will be made in the remainder of this article; Sommerhoff addresses this issue with the important concept of {\it a subject being in a state of expectancy} (\cite{som90}, page 67) for an event, but for our present discussion, as we are going to show,  this concept is not immediately relevant. Second, representations of the first and second kind do not contain representations of merely possible objects, events etc., as these are the elements of the subject's imagination. Sommerhoff argues that the latter category is certainly necessary for processes such as thought, but it is not a necessary condition for what we mean by being conscious about the world, the self-in-the-world and about events.  We adopt his opinion,  and hence for our definitions of (strong and weak) virtual reality we will exclude any direct reference to representations of objects of imagination. Third, for our preliminary understanding of virtual reality given in this section it is not necessary to involve specific assumptions about second order self-awareness. So, we presume here the existence of representations that register the occurrence of events, i.e. second order representations, without making this presumption explicit in our preliminary definition of virtual reality. This situation will change in Section \ref{sts}, when we will mathematically express representations of the first and the second category expressed through the modal language of the so-called Structural Theory of Sets, a hyperset theory.
 
The two characteristics of the first category, comprehension and coherence,  we can identify in terms of the two qualitative constituents of virtual reality, namely presence and interactivity. A comprehensive representation of the world and the self-in-the-world by a subject presupposes that this subject, who is a self at the same time, exhibits a sense of being in the world, but this is what presence means in the virtual reality context. On the other hand, interactivity in virtual reality requires for a  conscious subject that the action and its response carried out between the subject and its environment as well as the action and the response carried out between the subject and the self-in-the-environment develop what we call a sense-making relationship; for example, this sense-making interdependency can be attributed to subjective locality, causality and determinism.  In this sense {\it actio} and {\it reactio} must form a coherent relationship within the internal representation of the subject. Otherwise, coherence, and therefore interaction, would get lost if action and reaction would be uncorrelated in the internal representation of the beholder. We stress at this point that although comprehension and coherence will not admit an explicit mathematical meaning in our approach, imposed qualities like locality, causality and determinism may change this situation. This stems from the fact that the latter can be expressed in mathematical terms quite naturally. However, these arguments will be elaborated elsewhere.  The remaining quality of virtual reality, immersion, is implicitly present in the representation category (1), since this statement presupposes the existence of multi-sensory stimulus that sufficiently covers the sensory apparatus of the conscious subject in order to maintain the impression of (virtual) reality. Without such  sufficient coverage of senses, presence and interactivity would get lost, too. Thus we have made plausible that our simple model of reality $R_0$ equipped with the subject $S$ and the world $W_0$  may qualitatively be conceived a virtual reality. This is not surprising, for our own physical world, which can be seen as a realization of $R_0$, meets all three qualitative constituents of virtual reality at a high level. In summary, these considerations motivate a first attempt for a definition of virtual reality.   
\begin{definition}
\label{vr}
Let $R_1$ be a proper subclass of $R_0$ with $R_1 \cap S = S$. Then $\{R_1, S\}$ is a virtual reality for a conscious subject $S$ if $S$ forms representations of category (1) and of category (2) when  $W_0$ is replaced with $W_1 = R_1 \setminus S$.
\end{definition}
This definition implies that internal representations, realized as the subject's higher cognitive functions, naturally incorporate the defining qualities of virtual reality. It is therefore a reformulation of the defining qualities of virtual reality in terms of the defining qualities of conscious experience on the background of Sommerhoff's ideas. Yet our attempt still is preliminary because it fails to make any direct statement about the structure of conscious experience as a whole, and so it lacks immediate practical use when it comes to analyzing the {\it structural richness} of virtual reality.

\section{Events and strong virtual reality \label{strong}}

 The goal of this section is to give a more specific variant of the previous definition---one that determines the totality of events that a subject may experience in a virtual reality environment. To reach it, we first introduce the basic elements of conscious experience as sets. Let $S$ be again a conscious subject, then a representation of the first category, which we refer to as an {\it event} $E_0$, encodes two concurrent types of associations between elements of the world and the self-in-the-world. There is the association with a certain subclass $P \subset W_0$ of the world established through the internal representation of the world, and---through the omnipresent representation of the own self-in-the-world---there is also the association of $P$ with the self $S$. We represent both associations with one symbol $E_0(P,S)$, and further presume that an event $E_0(P,S)$ always forms a class for a given conscious subject $S$ and a given class $P$ in the world. Concurrently, second order self-awareness of an event $E_0(P,S)$, i.e. a representation of category (2), is given through the symbol $E^*_0(P,S)$. Thus $E^*_0(P,S)$ again is a class and it encodes the occurrence of an event $E_0(P,S)$, referred to as a second order event (Conversely, we may call events `first order events' but we will use the short form `event' throughout this work.). To illustrate this situation, imagine the event $E_0(P,S)$ of consciously looking at a physical object, then the object becomes the class $P$ here, the latter being a proper part of the subject's physical world, while $E^*_0(P,S)$ represents the second order event that registers the event of looking at the object.

It is then natural to postulate that every conscious subject $S$ has the ability to represent a whole collection of events, and we assign the symbol $V_0$ to this collection. Now we make the following assumption about the structure of $V_0$.\medskip

{\bf Assumption 1} {\it For a given reality $R_0$ and a conscious subject $S$ the collection  $V_0$ of all events is a universe for ZFC set theory.}
 \medskip

By a {\it universe for ZFC set theory} we mean a collection $V_0$ of sets that is a model $\mathcal{M}$ of ZFC set theory, i.e. a realization of the axioms in ZFC set theory by means of ordinary sets. Here, Assumption 1 becomes essential for our further considerations because it manifests the nature of the representations of the first category. This assumption also represents a specific reference to historical times in cultural history when virtual reality technology was  still undeveloped, as explained in the introduction section. But specifying the totality of events as a universe for ZFC set theory is still arbitrary at this stage, because we have not explained why events should be understood  as wellfounded sets in ZFC set theory.  We address this problem and justify Assumption 1 in Section \ref{empiric}, but for now we use it without further explanation as a reference point in our classification of virtual reality.

Let $\dom(V_0) \subseteq R_0$ be the domain of the universe $V_0$, that is the union of all classes $P \subset R_0$ that give rise to events $E_0(P,S)$. We choose  a subclass of $W_1 \subset \dom(V_0)$ such that $R_1 = W_1 \cup S$ is a virtual reality in the sense of Definition \ref{vr}, and thus we denote, in analogy to the collection $V_0$, as $V_1$ the collection of all events $E_1(P,S)$ with $P \subseteq W_1$. Equipped with this notation we are ready to introduce the concept of {\it strong virtual reality}.
\begin{definition}
\label{svr}
Let $R_1 \subset \dom(V_0) \cup S$ be a class with $R_1 \cap S = S$. Then the tuple $\{R_1, S, V_1\}$ is a strong virtual reality for a conscious subject $S$ if $\{R_1, S\}$ is a virtual reality with $V_1 \subseteq V_0$.
\end{definition}

This definition characterizes all those virtual realities whose collections of events are embedded within the universe $V_0$ for ZFC. Strong virtual reality may thus be conceived as a virtual reality that emerged from a part of the subject's original world with the addition that the events it mediates become consistent with the subject's universe $V_0$. In this sense, a subject $S$ would never experience a strong virtual reality as structurally {\it richer} than the subject's original universe $V_0$.  For example, we may think of strong virtual reality in terms of recent advances in technology where high-resolution displays, head-mounted-displays, headphones, data-gloves as well as properly designed software are used to process the information flux through the sensory interface \cite{vrinterface}; together they aim to conciliate presence, immersion and interactivity--although still at rather low quality. Hence, it is permissible to classify today's virtual reality technology as an early form of strong virtual reality, i.e. as a technology driven strong virtual reality that is now emerging in human culture.

We additionally remark that the above definition is general in that it does not specify the events to which it refers, except that it requires all events to be mapped onto wellfounded sets. Therefore two different virtual reality environments, i.e. two that present a very different content or that are based upon different technologies, may turn out to be quite similar, as sets or as classes, in strong virtual reality, because this definition neither does dictate technical design nor the presented content.

\section{Weak virtual reality \label{weak}}
\subsection{Weak virtual reality and Aczel's non-wellfounded set theory \label{weaksub}}

We give a more general statement to Definition \ref{svr}, which extends the collection of virtual reality created events.

\begin{definition}
\label{wvr}
Let $R_1 \subset \dom(V_0) \cup S$ be a class with $R_1 \cap S = S$. Then $\{R_1, \,^*V_1, S\}$ is a weak virtual reality for a conscious subject $S$ if $\{R_1, S\}$ is a virtual reality, and if $V_0 \subset \,^*V_1$ for any universe $V_0$ for ZFC set theory.
\end{definition}

In contrast to strong virtual reality, a realized weak virtual reality would have dramatic consequences for a conscious subject $S$, because the collection of conscious events formed from this kind of virtual reality would be a richer structure than the subject's original universe, or, in logical terms: $^*V_1$ strongly implies $V_0$. Hence, in this environment the subject could experience something that might be by intuition called a {\it reality shift} from $V_0$ to $^*V_1$. This argumentation, however, is insufficient because we cannot know yet in what sense events are associated with non-wellfounded sets at all, and to what extent the mathematical difference between wellfounded and non-wellfounded sets should result in a corresponding difference of the subject's cognitive experience. Both problems will be addressed throughout the remainder of this work.

Before addressing these problems, we want to examine the question whether this definition is mathematically meaningful at all. And, indeed, non-well\-founded set theory suggests that the concept of weak virtual reality is not meaningless, because a universe for non-wellfounded set theory exists which already contains any universe for ZFC. We follow a well-known and intuitive approach to non-wellfounded set theory \cite{acz88, dev92}, and introduce the so-called Anti-Foundation-Axiom (AFA) which will replace the Axiom of Foundation in ZFC set theory, i.e the axiom saying that the relation $\in$ is wellfounded.  The resulting non-wellfounded set theory establishes a one-to-one correspondence between graphs (the pointed, directed graphs in fact) and sets via AFA. It therefore can be seen as the source of a logically weaker and thus mathematically broader concept of set. For that reason that we use the term {\it weak} virtual reality.

Let $G = \{N, E\}$ be a directed graph, where $N$ is the set of nodes and $E$ is the set of edges all being ordered pairs of nodes. For $(x,y) \in E$ we write $ x\rightarrow y$; in that case $x$ is a parent of $y$ and $y$ is a child of $x$. A path in $G$ is a finite or infinite sequence of nodes, each of which (except the first) is a child of its predecessor. If there is a path from node $x_1$ to node $x_k$, we say that $x_k$ is an descendant of $x_1$. A directed graph is said to be pointed if there is a unique node $x_0$ (the root of the graph) such that all other nodes are descendants of $x_0$. From now on we always mean a directed, pointed graph when we simply write `graph'. The Anti-Foundation Axiom then reads as: {\it Every graph depicts exactly one set (AFA).} This axiom gives rise to the existence of non-wellfounded sets, such as sets formed from graphs having loops of edges (circular sets) or from graphs with an infinite number of edges in a path, and these 'sets' cannot be constructed in ZFC set theory.  If we replace the Axiom of Foundation with the Anti-Foundation Axiom in ZFC set theory we obtain a collection of axioms which we denote as ZFC$^-$+AFA set theory. The immediate question then is whether the set theory ZFC$^-$+AFA is relatively consistent with regard to the original ZFC set theory, i.e. if there is a model of $\mathcal{M}$ of ZFC set theory that is a submodel of ZFC$^-$+AFA set theory. The next Theorem makes a clear statement about that.

\begin{theorem}
\label{arc}
{\sc{ [Aczel's relative consistency result]}} If  $\,\,V'$ is a universe for ZFC set theory, then there is a universe $V^*$ for ZFC$^-$+AFA such that $V' \subset V^*$.
\end{theorem}

The proof, originally given in \cite{acz88}, is guided by two questions: ``When are two sets pictured by the same graph?'' and ``When do two graphs picture the same set?''. We want to sketch the main points of the proof. A {\it system} is a generalization of the concept of a graph in the sense that the collections of nodes and directed edges may now be proper classes also. Any system $M$ is required to satisfy the requirement that, for each node $x$, the collection $\ch_M(x) =\{x'\in x | x \rightarrow x'\}$ of all children is a set. Clearly, any graph is a system. A {\it system map} $\pi:M \rightarrow M'$ between two systems $M$ and $M'$ is a map such that for all $x \in M$, $\pi$ maps the children of $x$ in $M$ onto the children of $\pi(x)$ in $M'$; i.e., for all $x \in M$ it is $\ch_{M'}(\pi(x)) = \{\pi(y)|y \in \ch_M(x)\}$.
As its main point, the proof of Theorem \ref{wvr} provides a canonical surjective system map $\pi: V \rightarrow V_c$, where $V$ is any system consistent with ZF set theory without the Axiom of Foundation (which we denote as ZF$^-$), and where $V_c$ is a class which turns out to be a {\it complete system}.  Given a system $M$, an $M$-decoration of a graph $G$ is just a system map $\pi_d:G \rightarrow M$. A complete system is a system $M$ such that every graph has a unique $M$-decoration. Since it then can be shown that every complete system is a model of ZFC$^-$ as well as that every complete system is a model of AFA at the same time, it follows that the system map $\pi$ establishes a canonical embedding of $V'$ in $V^*$, which completes the proof.

Bringing back our attention to the notion of weak virtual reality, we thus have the following result immediately.
\begin{corollary}
Let $\{R_1, S, V_1\}$ be a strong virtual reality. Then there is a canonical system map $\pi:V_1 \rightarrow\, ^*V_1$ such that $V_1 \subseteq V_0 \subset\, ^*V_1$. Moreover, $^*V_1$ is a universe for ZFC$^-$+AFA set theory.
\end{corollary}
For any given strong virtual reality, there is a larger universe for ZFC$^-$+AFA that is relatively consistent to $V_1$, i.e. $V_1 \subseteq V_0 \subset\, ^*V_1$, which indeed suggests that Definition \ref{wvr} is mathematically meaningful. Then, as already indicated, the next task is to demonstrate whether or not $^*V_1$ allows for a collection of events, i.e. whether or not the elements of $^*V_1$ are representations of category (1) and category (2) for a conscious subject $S$. However, this question cannot be answered satisfactory unless additional information is provided about the structure of the universes $V_1$ and $^*V_1$ along with their members, i.e. their respective events. Such information especially concerns a suitable mathematical representation of the {\it internal state} of the conscious subject $S$, and further, how events may change the internal state. The notion `change' implies that there is a temporal metaphor within, and thus one may ask whether some process governs the subject's state and possibly causes a transition from $V_1$ to $^*V_1$, i.e. a transition from strong into weak virtual reality.

\subsection{Heuristic arguments \label{empiric}}

Having introduced strong virtual reality and weak virtual reality as rather formal entities, our next aim is to discuss the evidence that both are relevant and practical concepts for studying the relationship between human cognition and virtual reality technology. We want to bring forward the idea that in weak virtual reality physical objects may be represented as non-wellfounded structures---something that controverts our everyday understanding of physical matter. Conversely, the aforementioned transition from strong to weak virtual reality could be accomplished by a technical realization of events that are associated with non-wellfounded objects in the virtual reality world. We begin with a general outline of this idea and discuss an example thereafter.

Our physical world consists of material objects that find a scientifically valid description among physical theories and mathematical structures consistent with our physical experience. And physical experience itself is gained through scientific experiments and through our everyday life in the world of physical phenomena. In a common understanding, physical objects are spatially limited structures in a three-dimensional geometrical space, i.e. in the physical space. Additionally, these objects are equipped with certain measures such as mass, charge or angular momentum.\footnote{Such measures exist in classical and in quantum physics.} Since this understanding implies that physical space is a topological space to which physical objects belong as sets, they must comply with an often unexpressed or even ignored condition: No physical object can be a proper part of itself. Clearly, this means that we cannot split an extended and material physical object into several parts, so that the latter become physical objects again, and then realize that one part is identical with the original object before the split. All physical objects are expected to follow this condition for otherwise contradictions would occur even at the elementary level of set theory. And from a certain standpoint this concern is justified, for in wellfounded set theories, like in set theories based on the Zermelo-Fraenkel axioms, any set $a$ having a proper subset $b \subset a$ such that $\neg (a \subset b)$ with $a = b$ must be disregarded, because any two sets $a$ and $b$ are equal if their members are equal (according to the Axiom of Extensionality) from which follows $a = a$, a contradiction to $\neg (a \subset a)$. Since set theory is seen as the basic mathematical theory describing the structure of collections, and so it can be utilized to represent {\it any} collection of physical objects, contradictions at this elementary level are to be avoided right away.\footnote{Moreover, any violation of the condition that no physical object can be a part or a member of itself would immediately dissent mass and energy conservation, for example.} At the same time, however, a contradictory situation can be the source of a deeper understanding for it prompts us to scrutinize our basic assumptions.

Let us briefly recall the current role of wellfounded set theory in the physical sciences. During the history of modern physics, set theory has largely been left unnoticed as a remote field in pure mathematics, and set theory itself has implicitly been equated with wellfounded (ZFC) set theory. Non-wellfounded set theories, although formulated rigorously as early as 1926 \cite{fin1924}, have not received any continuous attention in the physical sciences. But such a restriction to wellfounded sets has concurrently placed the physical sciences in a certain situation or, differently put, within a certain context. Within such context, the above construct $a$ cannot be a mathematical structure `set', and therefore it cannot function as a scientifically valid model of a physical object either. Again, this tells us that in the chosen context of wellfounded set theories and notably on the background of Zermelo-Fraenkel set theory no physical object can be a proper part of itself if we want to anticipate obvious contradictions. This observation brings out a reason why modern physical theories so far have limited themselves to mathematical structures like sets, groups, algebras, etc., that without exception are consistent with ZFC set theory. The reason is that  non-wellfounded set theories have not been considered physical. Consequently, a universe $V_0$ for ZFC set theory containing wellfounded sets only appears to be a legitimate description for the entirety of first order internal representations of the physical world {\it in this context}. Indeed, this argument recovers our statement given in Assumption 1. It thus characterizes the current situation in which all physical objects are expected to be wellfounded.  But is this situation necessarily permanent?  

\begin{figure}
\begin{center}
\includegraphics [width=7.5cm]{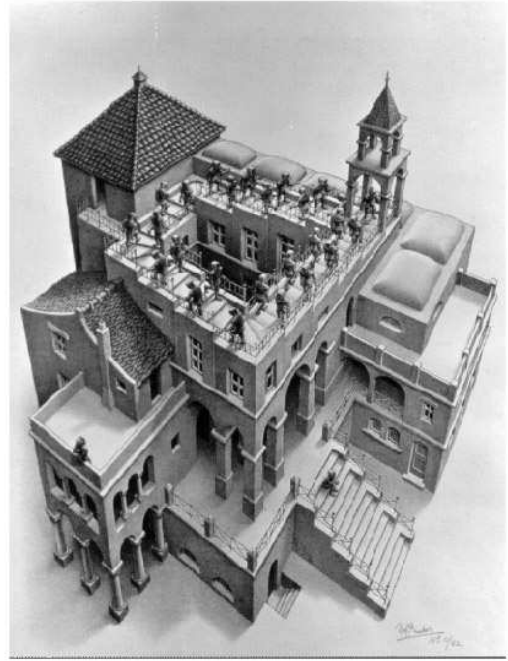}
\end{center}
\caption{M. C. Escher's lithograph ``Ascending and Descending'' (1960). The pictured staircase on the top of the building is  often characterized as `impossible' because of its circular structure. In weak virtual reality such construction is conceivable as part of a subject's world. Copyright M.C. Escher Foundation.  \label{escher}}
\end{figure}

To approach this question, we use an example given in Fig. \ref{escher}. It shows the famous lithograph of M.C. Escher titled ``Ascending and Descending''. As many of Escher's graphical works it pictures a situation considered {\it impossible} in the physical world. In this picture, a staircase is shown connecting four levels in a building. To follow the stairs represented in the scene, we may begin our way from the floor located at the left-hand-side of the picture--it is the corner with a two-story tower. From here a descent is possible, i.e. a way downstairs towards the ground. After making three more such descends, and after passing three more floors, something unfamiliar happens. Even though a way down has always been chosen, and so an arrival at a lower level somewhere beneath the starting point has been expected, the way has brought us back to the little tower---our starting point. But this must be impossible in physical space if the latter is understood as three-dimensional Euclidean space and if the embedded building is considered a rigid material object. We may illustrate this situation using wellfounded set theory. For this purpose we represent each of the four floors by members of the set $\{s_0,s_1,s_2,s_3\}$, and further assume that each member is a set $s_i = \{s_{i-1}\}$ with $i \in \mathbb{Z}/4 \mathbb{Z}$ (i.e., $i$ is an element of the cyclic group of order 4), so that the process of descending  from, say, floor `$s_1 = \{s_0\}$' to floor `$s_0$' is the membership relation between $s_1$ and $s_0$, viz. $s_1 \ni s_0$. Hence, by making four consecutive descents we obtain the nested structure $s_3\ni s_2 \ni s_1  \ni s_0 \ni s_3$, which---in contradiction to our assumption---cannot represent a set $s_3$ in wellfounded set theory because it implies an infinite hierarchy of membership, i.e. a non-wellfounded structure. This example again shows that in general geometric self-reference and circularity as aspects of physical objects cannot be described within wellfounded set theory.

From a mathematical point of view, events associated with non-well\-founded objects do not cause difficulties if we allow to shift our context from wellfounded to non-wellfounded set theory. For example, in ZFC$^-$+AFA set theory we can resolve the contradiction caused by introducing self-referential structures as sets.  In the above case, the structure $s_3\ni s_2 \ni s_1  \ni s_0 \ni s_3$ clearly depicts a non-wellfounded set in ZFC$^-$+AFA.

We see that despite the problems it causes in wellfounded set theory, Escher's picture gives a very clear impression of the represented scene, with high level of detail and with technical finesse---qualities that make it familiar and concrete despite its abstract strangeness. And from this standpoint we may go one step further and suggest an experiment, in which this special scene is presented to a human subject immersed in an virtual reality environment. The experiment would document the subject's response to this situation, and two distinct outcomes are conceivable. Either a negative result, i.e. the subject would not significantly develop presence, immersion and interactivity with the presented environment and would thus perceive the visual stimulus simply a plain picture representing a fictitious scene (Fig. \ref{escher}). Or a positive result, upon which the subject would experience presence, immersion and interactivity at a level where an internal representation of the staircase in Escher's ``Ascending and Descending'' would become an event associated with a physical object in the subject's world.\footnote{We do not require such an experiment to be conducted using exclusively one of Escher's works, of course.} We stress, however, that for such a  level to be reached virtual reality technology would have to be fairly advanced, i.e. a certain technological standard would have to become available in order to conduct our hypothetical experiment. We think that computer technology may provide such a standard, and first attempts exist that put `impossible objects' into computer generated virtual reality environments. For example, in the project ``Escher revisited in VR valley'' \cite{escher} several of Escher's graphical works have been transformed into animated computer graphics. Continuous spatial motion in three dimensions and sophisticated light effects have been added to the original scenes, and, in comparison to Escher's originals, the resulting virtual reality quality of the computer generated scenes has increased. Given this progress, we expect that technological evolution will continue to refine simulations of `impossible' objects and scenes that include non-wellfounded structures. This advancement may eventually lead to a positive result of our suggested experiment, documenting the subject's ability to conceive new and unprecedented events associated with a non-wellfounded sets represented in virtual reality environments. Evidently, such result would document the subjective transition into weak virtual reality.

\section{Structural Theory of Sets as a general approach and concluding remarks\label{sts}}

In this work,  we began our investigation of  the relationship between consciousness, virtual reality and hypersets by formulating the defining qualities of virtual reality in terms of Sommerhoff's first and second order self-aware\-ness. We have then taken this formulation to define strong and weak virtual reality, and gave initial theoretical and heuristic arguments for why we think that weak virtual reality can be created with non-wellfounded objects realized within virtual reality environments. We now return to our initial point by showing that in turn  non-wellfounded set theory can devise first and second order self-awareness in pure logical terms.  Also, we are going to demonstrate that strong and weak virtual reality admit a representation within a general non-wellfounded set theory, the Structural Set Theory, which contains Aczel's non-wellfounded set theory as a special case. This step indicates that consciousness, weak virtual reality and hypersets can be put into one coherent theoretical framework.

Baltag has developed a general axiomatic set theory, the Structural Theory of Sets--STS \cite{bal99}, formulated in sentences of modal logic, i.e. in terms of a modal language generated by three logical rules (see Appendix section): negation, conjunction and unfolding. Modal logic itself is a generalization of classical binary logic as the former introduces two propositional modalities, referred to as necessity and possibility, whereas the latter does not make such a distinction and treats all valid propositions as necessary. The connection between modal logic and sets is established through so-called {\it satisfaction axioms}, which guarantee that under reasonable conditions a modal sentence is represented (satisfied) by a set.  Hence modal sentences are seen as statements about sets that can be satisfied by sets, and hypersets in particular are satisfied by sentences including infinite logical conjunctions. 

One result of STS is the existence of a {\it universal set} $ U$, which contains all sets satisfied by valid modal descriptions, and $th(U)$, called the {\it modal theory of} $U$, denotes the collection of these descriptions. Since the universal set has members that are wellfounded and non-wellfound sets, we propose $U$ for entirety of events of a conscious subject $S$ so that $th(U)$ becomes the theory of $U$ for the subject $S$, i.e. $th_S(U)$. As a consequence,  STS offers here a direct connection to first and second order events: Suppose $E$ is a set in $U$, satisfied by a modal description $\varphi \in th_S(U)$, and associated with an event of first order self-awareness. Then the unfolding rule says that there is another valid sentence, $\Diamond\varphi \in th_S(U)$, now capturing the information that $U$ has a member described by $\varphi$. In turn, the description $\Diamond\varphi$ is satisfied by a set $E^*$, yet another member of the universal set $U$, and so the set $E^*$ can be identified as a (second order) event associated with second order self-awareness. This observation justifies our representation of events through sets as given in Section \ref{strong}, and it can be summarized in the following proposition (A more detailed outline of our argument together with a brief introduction to STS is given in the Appendix.)   

\begin{proposition}
\label{eventsts}
(a) Sommerhoff's terms of first and second order self-aware\-ness can be represented in modal language. (b) First and second order events in strong and in weak virtual reality can be represented in STS as sets that satisfy consistent (infinitary) modal sentences in $th_S(U)$.  
\end{proposition}

\medskip

We have claimed that set theory provides a reasonable framework for studying strong {\it and} weak virtual reality. This claim is now further supported by the above proposition which specifies a mathematical form for the entirety of conscious events, including first and second order self-awareness, in one non-wellfounded set: the universal set $U$ in STS. 

Yet several unexplored directions remain, and one is the problem of identifying those modal descriptions which would result in coherent internal representations of the world and the self-in-the-world. As we have argued, only these peculiar events mediate presence, immersion and interactivity, and there is no reason to believe why all modal descriptions should lead to conscious events. We have briefly mentioned in Section \ref{reality} that locality, causality and determinism may be necessary qualities, but at this stage it not quite obvious how to integrate these qualities into our proposed framework. 

Another direction regards the potential influence of virtual reality technology on the physical sciences. Since virtual reality devices may not only have the capacity to simulate the original physical world in strong virtual reality, but also to transcend it in the context of weak virtual reality, critical questions about the scientific character of physical experience in the light of virtual reality should follow. For example, is it imaginable that physical laws, i.e. the basic conditions of physical phenomena as currently known in science, will at some point become largely arbitrary as virtual reality technology---a technology controlled by humans---develops further? It appears that we cannot analyze such a provocative question with scientific scrutiny until we have explored further the scientific conditions and implications of conscious experience.  

\appendix

\section{Appendix}
\label{App}

We employ a generalization of ZFC$^-$+AFA set theory and follow the original work of Baltag \cite{bal99}, in which a {\it structural} concept of sets is introduced, and we briefly outline some of the elementary ideas in STS. A structural understanding of sets is in a sense {\it dual} to the classical iterative (i.e., synthetic) concept of set. While in the latter we consider sets as built from some previously given objects in successive stages (the {\it iterative concept of set}), the former presupposes that {\it a priori} a set is a {\it unified totality} that reveals its abstract membership structure only step by step through the process of structural unfolding. This stepwise discovery of the set structure is generated by imposing questions (which Baltag calls analytical experiments) to the initial object; the answers to these questions are the stages of structural unfolding.  Thus at each stage we have a partial description of the object considered. The unfolding process can be defined by recursion on the ordinals: for every ordinal $\alpha$ and every set $a$, the {\it unfolding of rank $\alpha$} is the set $a^\alpha$, given by:
\begin{eqnarray}
\nonumber
a^{\alpha +1} &=& \{b^\alpha : b \in a\}\\
\nonumber
a^\lambda &=& \langle a^\alpha \rangle_{\alpha < \lambda} \, \mbox{, for limit ordinals,} \, \lambda
\end{eqnarray}
Surely, this definition is meaningful for all wellfounded sets, but for a larger universe it is inappropriate in general since $\in$-recursion is equivalent to the Axiom of Foundation.

To find a definition of structural unfolding for more general objects, i.e. systems or classes,  Baltag takes seriously the fact that at every ordinal stage we can only have a partial description of a system. An essential ingredient here is the {\it observational equivalence between systems} -- a generalization of the bisimulation concept for systems \cite{acz88, dev92}. In fact, bisimulations do not have an answer to the question ``When do two systems (sets) depict the same set (system)?'' whenever large systems are considered, i.e. whenever those systems fail to be graphs. On the contrary, observational equivalence is given by modal equivalence, and not by the usual definition of bisimulation, and it turns out that with {\it infinitary modal logic} observational equivalence between systems can be defined to incorporate even large systems or classes. 

In STS,  a modal theory $th(a)$ for every set or class  $a$ is given through the so-called satisfaction axioms, and before we quote these axioms we may first introduce the underlying modal language.
\begin{enumerate}
\item Negation. Given a possible description $\varphi$ and an object $a$, we construct a new description $\neg \varphi$, to capture the information that $\varphi$ does not describe $a$.
\item Conjunction. Given a set $\Phi$ of descriptions of the object $a$, we accumulate all descriptions in $\Phi$ by forming their conjunction $\bigwedge\Phi$.
\item Unfolding. Given a description $\varphi$ of some member (or members)of a set $a$, we {\it unfold} the set by constructing a description $\Diamond\varphi$, which captures the information that $a$ has some member described by $\varphi$.
\end{enumerate}
The language generated by these three rules is called {\it infinitary modal logic} which allows infinite conjunctions. With $\bigvee$ and $\square$ as the duals (obtained by substituting $\wedge \mapsto \vee$ and $\Diamond \mapsto \square$) to $\bigwedge$ and $\Diamond$, respectively, we can introduce the following other operators:
\begin{eqnarray}
\nonumber
\Diamond \Phi &=:& \{\Diamond \varphi: \varphi \in \Phi\}\,,\\
\nonumber
\Box\Phi &=:& \{\Box\varphi : \varphi \in \Phi\}\,,\\
\nonumber
\varphi \wedge \psi &=:& \bigwedge\{\varphi, \psi\}\,,\\
\nonumber
\varphi\vee\psi &=:& \bigvee\{\varphi, \psi\}\,,\\
\nonumber
\bigtriangleup\Phi &=:& \bigwedge\Diamond\Phi \wedge \Box \bigvee\Phi\,.
\end{eqnarray}
The satisfaction axioms presume the existence of a class a $Sat$, each element of $Sat$ is a pair of a set $a$ and a modal sentence $\varphi$. Writing $a \models \varphi$ for $(a, \varphi) \in Sat$, these axioms read as
\begin{eqnarray}
\nonumber
&{\rm{(SA1)}}& a \models \neg \varphi \quad {\rm{iff}} \quad a \not\models \varphi\\
\nonumber
&{\rm{(SA2)}}& a \models \bigwedge \Phi \quad {\rm{iff}} \quad a' \models \varphi\quad {\rm{ for \,\, all}} \quad \varphi \in \Phi\\
\nonumber
&{\rm{(SA3)}}&  a \models \Diamond\varphi \quad {\rm{iff}} \quad a' \models \varphi\quad {\rm{ for \,\, some}} \quad a'\in a
\end{eqnarray}       
With this setting the notion of unfolding of a set $a$ admits now an expression through modal sentences $\varphi^\alpha_a$ defined for any cardinal $\alpha$ as
\begin{eqnarray}
\nonumber
\varphi^{\alpha+1}_a &=:& \bigtriangleup\{\varphi^\alpha_b : b \in a\}\,,\\
\nonumber
\varphi^\lambda_a &=:& \bigwedge\{\varphi_a^\beta:\beta<\alpha\}\quad {\rm{ for \,\, limit \,\, cardinals}}, \lambda\,.
\end{eqnarray}
Unfoldings of rank $\alpha$ are {\it maximal} from an informational point of view as they gather all the information that is available at stage $\alpha$ about a set and its members. In formal language this statement reads as the proposition: $b \models \varphi^\alpha_a$ iff $b^\alpha = a^\alpha$. This explains the notion of {\it observationally equivalent}: two sets, classes or systems are said to be observationally equivalent if they satisfy the same infinitary modal sentences, i.e. if they are modally equivalent. In STS, the existence of sets is guaranteed by the bijection $th(\cdot)$ between maximal weakly consistent theories and the sets (This correspondence is an immediate consequence of the {\it Super-Antifoundation Axiom in STS}.). Weakly consistent theories are those theories in which all sub-collections of descriptions that are satisfied by a set are closed under infinitary conjunctions. It follows that non-wellfounded sets or classes are exactly those which do not admit satisfaction by any finite conjunction in infinitary modal logic.\medskip

On this background, we now make the observation that STS---constructed on the basis of infinitary modal logic--- provides a tool to reformulate our description of Sommerhoff's first order and second order self-awareness. First of all, a {\it universal set} $U$ exists in STS being observationally equivalent to the universal class $U_c = \{x: x \mbox{ is a set}\}$. We want to use $U$ as the collection of all events $E_1$ in weak virtual reality. This is possible because it can be shown that any universe $^*V_1$ for ZFC$^-$+AFA set theory is observationally equivalent to the set $U$ \cite{bal99}. This suggests that  a weak virtual reality $\{R_1, U, S \}$ can be defined within STS as well.  To see this, consider any arbitrary event $E_1(P,S) \in U$ associated with its modal description $\varphi \in th(U)$, i.e. $E_1(P,S) \models \varphi$, then events themselves may be understood as modal descriptions mediating first order self-awareness. 

The theory $th(U)$ is formulated in the infinitary modal language of the subject $S$, thus we write $th_S(U) = th(V_0)$ to account for this dependence on the presence of a conscious subject $S$. Any event $E_1(P,S) \in U$ is then understood as the image of $P \subseteq W_1$ under the map $d := th_S \circ th^{-1}_S$, i.e. $E_1(P,S) = d(P) = th_S^{-1}(th_S(P))$, with $th_S$ being the operator mapping classes $P \subseteq W_1$ onto their modal theories and $th^{-1}_S$ is the inverse map; both maps exist in STS \cite{bal99} and together they define the {\it denotation} function $d := th \circ th^{-1}$ with $d: R_1 \rightarrow U$.  In that manner first order self-awareness are mathematically represented by the denotation $d$. And second order self-awareness is established here concurrently through the unfolding rule, i.e. rule (3), as $\Diamond\varphi$ now encodes the information that $U$ has a member described by $\varphi$. But $\Diamond \varphi$ as a sentence is a member of $th_S(U)$, because $th_S(U)$ already contains the collection $\{\Diamond \varphi: \varphi \mbox{ is a consistent modal sentence}\}$ \cite{bal99}. Then $th^{-1}_S(\Diamond \varphi)$ must be a member of $U$, and we identify it as $th^{-1}_S(\Diamond \varphi) = E_1^*(P,S)$. The latter statement clearly resembles second order self-awareness as described in Section \ref{svr}. Finally, since rule (1) and rule (2) are satisfied, i.e. they recognize the fact that modal descriptions of events can be logically combined to create valid descriptions of sets again, Proposition \ref{eventsts} follows.

We note that the special case of strong virtual reality is obtained by using a denotation $d$ which maps classes in $R_1$ onto wellfounded sets. These sets  altogether comprise the totality of events in strong virtual reality $V_1$. In that case the corresponding theory is $th_S(V_1) \subset th_S(U)$.


\end{document}